\documentclass[pre,aps,twocolumn,superscriptaddress]{revtex4-2}

\usepackage[latin1]{inputenc}
\usepackage{graphicx}
\usepackage{amsmath}
\usepackage{amssymb}
\usepackage{color}
\usepackage[dvips]{epsfig}
\usepackage{epstopdf}
\usepackage{times,mathptmx}
\usepackage{epstopdf}
\usepackage{natbib}

\DeclareMathOperator{\arsinh}{arsinh}

\DeclareMathOperator{\Ei}{Ei}

\begin{document}
\title{Electrophoresis of ions and electrolyte conductivity: from bulk to nanochannels}
\author{Olga I. Vinogradova}
\email[Corresponding author: ]{oivinograd@yahoo.com}
\affiliation{Frumkin Institute of Physical Chemistry and
Electrochemistry, Russian Academy of Science, 31 Leninsky Prospect,
119071 Moscow, Russia}
\author{Elena F. Silkina}
\affiliation{Frumkin Institute of Physical Chemistry and
Electrochemistry, Russian Academy of Science, 31 Leninsky Prospect,
119071 Moscow, Russia}

\date{\today }

\begin{abstract}
When electrolyte solutions are confined in micro- and nanochannels their conductivity is significantly different from those in a bulk phase. Here we revisit the theory  of  this phenomenon by focusing attention on the reduction in the ion mobility with the concentration of salt and a consequent impact to the  conductivity of a monovalent solution, from bulk to confined in a narrow slit. We first give a systematic treatment of electrophoresis of ions and obtain equations for their zeta potentials and mobilities. The latter are then used to obtain a simple expression for a bulk conductivity, which is valid in a concentration range up to a few molars and more accurate than prior analytic theories. By extending the formalism to the electrolyte solution in the charged channel the equations describing the conductivity in different modes  are presented. They can be regarded as a generalization of prior work on the channel conductivity to a more realistic case of a nonzero reduction of the zeta potential and electrophoretic mobility of ions with salt concentration. Our analysis provides a framework for interpreting
measurements on the  conductivity of electrolyte solutions in the bulk and in narrow channels.

\end{abstract}

\maketitle

\section{Introduction}\label{sec:intro}

The conductivity of strong electrolyte solutions  confined in narrow channels is a subject that currently attracts much experimental and  theoretical  research  effort~\cite{schoch.rb:2008,bocquet.l:2010,vinogradova.oi:2023}. One  topic of  interest  concerns  the nature of  the  ion transport in micro- and nanofluidic  channels and,
in particular, how this differs from that of a bulk ion transfer. The best known example is probably that of a dilute  electrolyte  solution becomes ``superconducting'' in a narrow channel, that is its conductivity $K$ may exceed the bulk one $K_{\infty}$  by orders of magnitude~\cite{stein.d:2004,schoch.rb:2005,vanderHeyden.fhj:2007}. That the electrolyte conductivity in the channel should be different from that of a bulk solution becomes apparent when one recognizes that there are two contributions to the increase in conductivity. These are an enhanced electrophoretic migration of ions relative to a solvent due to their enrichment in the charged channel and a convective electro-osmotic ion transfer  arising in the channel subject an applied tangential electric field $E$. The former contribution is only present for finite thickness $H$ whereas the latter is present in semi-infinite systems ($H \to \infty$).


During the last decade extensive efforts have gone into theoretical investigating the dependence of $K$ on electrostatic boundary conditions at the channel walls, their wettability, and electrolyte concentration~\cite{ren.y:2008,vinogradova.oi:2021,green.y:2022}.
 Theory has made significant advances leading to interpretation of experiments on conductivity in a wide variety of channels. Such experimental observations as the conductivity plateau in dilute solutions, power-law scaling and variation of its exponents depending on surface state and salt concentration have been quantitatively interpreted~\cite{shan.yp:2013,balme.s:2015,secchi.e:2016,biesheuvel.pm:2016}. However, we gained the impression that several aspects of the an electrophoretic migration of bulk and confined ions have still been given insufficient attention.

The velocity of migrating relative to a fluid ions is linear in $E$, and the proportionality coefficient is referred to as an electrophoretic mobility.
The simplest expression, and the one that makes a connection of a mobility with ion hydrodynamic radius, can be derived by postulating the Stokes resistance to the  ion electrophoretic propulsion. If we denote as $m$ the mobility of cations, in the case of 1:1 electrolyte this expression  reads
\begin{equation}\label{eq:inf}
m = m_0 = \dfrac{e }{6\pi \eta R},
\end{equation}
where $R$ is the hydrodynamic radius of the cation. The mobility of univalent anions of the same radius is then equal to $-m$. The current density generated in a solution of number density $n_{\infty}$ is then $J = 2 e E n_{\infty}m$. Using $J = K_{\infty}E$ (Ohm's law), where $K_{\infty}$ is the electrical conductivity of a bulk solution, and substituting \eqref{eq:inf} yields
\begin{equation}\label{eq:Kinf}
 K_{\infty} = K_{\infty}^0 = \dfrac{e^2 n_{\infty }}{3 \pi \eta R},
\end{equation}
which  in electrochemistry textbooks is termed the conductivity at an \emph{infinite dilution}.
Since the ion mobility given by Eq.\eqref{eq:inf} does not depend on salt, the bulk conductivity in this model  increases linearly with the concentration, $K_{\infty}^0 \propto n_{\infty}$.

While  Eqs.\eqref{eq:inf} and \eqref{eq:Kinf} are traditionally invoked in the interpretation of the conductivity data in the micro- and nanofluidic channels, even when the electrolyte concentration is quite high, their origin in terms of electrophoretic migration of ions remains obscure. It is well known that the resistance to electrophoresis of an ion is greater than the Stokes drag to the motion of an uncharged particle. Thus hydrodynamic arguments, based on the Stokes force, may  become unreliable when the electrolyte concentrations are high enough.
Indeed, experiments on bulk conductivity have shown that it is generally smaller than predicted by \eqref{eq:Kinf} and reduces with increasing $n_{\infty}$. Kohlrausch found that the decrement in mobility is proportional to the square root of concentration, which gives the following limiting formula referred nowadays to as  Kohlrausch's empirical law~\cite{kohlrausch.f:1900}
\begin{equation}\label{eq:Kolr}
  K_{\infty} \simeq  K_{\infty}^0 - \alpha   n_{\infty }^{3/2},
\end{equation}
where $\alpha$ is some constant, which is often determined from the conductivity measurements themselves. \citet{onsager.l:1927} appears to have been the first to
justify the form of Eq.\eqref{eq:Kolr} theoretically, to evaluate the constant $\alpha$, and to  argue that this coefficient of the square root term does not include the ionic hydrodynamic radius $R$. According to the derivation, Eq.\eqref{eq:Kolr}  should be applied only at the \emph{very high dilution}, but will become less accurate or invalid for larger concentrations. It would be of considerable interest to determine the regime of validity of this asymptotic equation for $K_{\infty}$. The same, of course, concerns $K_{\infty}^0$. In essence, the  terms very high and infinite in  context of dilutions  have  been not yet  rigorously defined, although there is a large amount of literature describing attempts to provide a more satisfactory theory by including into \eqref{eq:Kolr} the higher order corrections. We refer the reader to the comprehensive review on bulk conductivity studies with the summary of these theoretical models that apply at a few millimolars~\cite{barthel.j:1968}.
 Later attempts at extension the theory to a molar concentrations have employed density functional approach, in several versions~\cite{chandra.a:1999,banerjee.p:2019,avni.y:2022}. This approach always involves many parameters and relies on numerical calculations. To the best of our knowledge, only  recently a relatively simple analytical approximation~\cite{avni.y:2022}.

Since the reduced conductivity is associated with the decrease in the ion mobility, the mobility $m$ that leads to Eq.\eqref{eq:Kolr} should be smaller than given by \eqref{eq:inf} and sensitive to salt. However, a quantitative interpretation of this effect from the point of view of modern theories of electrophoresis has not  been given yet. Given the current upsurge of interest in the conductivity of electrolytes in narrow channels and its numerous applications
it would seem timely to revisit the issue of the ion mobility and to bring some more advanced knowledge on electrophoresis to bear on the problem of a bulk  conductivity, and its repercussions for a conductivity in channels.

In this paper we present a derivation of simple expressions for an electrophoretic mobility of ions and a consequent bulk conductivity  valid in quite large concentration range, namely, up to a few mol/l. In limiting situations of infinite and very high dilutions our compact conductivity equation reduces to Eqs.\eqref{eq:Kinf} and  \eqref{eq:Kolr}, allowing us to clarify their status and to specify the range of applicability. At molar concentrations our simple formula appears to provide a much better fit to the experimental data than prior models.  These results allowed us to revisit the hot issue of the electrolyte  conductivity in the channels, now by taking into account the effect of reduction of zeta potentials (mobilities) of ions with salt.  We are unaware of any previous work that has even mentioned this effect: it has previously been completely ignored.
 We shall see that the extension of our theory of a bulk conductivity to the channel case provides new insight into the physics of  migration and convective transfer of ions, concentration scaling of the conductivity, and the nature of conductivity enhancement in micro- and nanochannels.

Our paper is arranged as follows. In Sec.~\ref{sec:model} we introduce the concepts of ion zeta potentials and provide calculations of their values as a function of salt concentration. The equations for an ion mobility and a conductivity expressed via zeta potentials of ions are then derived. They recover Eqs.\eqref{eq:Kinf} and \eqref{eq:Kolr} in the infinite and very high dilution situations that are now well defined and provide a rather accurate description at low dilution.
Section~\ref{sec:channel} discusses a conductivity of a single charged channel in contact with a bulk electrolyte reservoir. Our main focus is on the effect of salt-dependent ion zeta potentials on the conductivity curves. The procedure for calculating the conductivity in the thick and thin channel modes is described, and derivations of analytic equations for $K$ are given. We argue that equations obtained for an infinite dilution can be employed safely to
interpret the results on a huge channel conductivity enhancement  at low bulk concentrations of salt. For the case of higher bulk concentrations however, the decrease in zeta potential significantly affects the $K$-curves and cannot be neglected.
 We conclude in Sec.~\ref{sec:concl}
with a discussion of open problems and possible future extensions of our results for more complex systems. In Appendix~\ref{a:1} we compare calculations from our theory with experimental data and predictions of some early theoretical models.

   \section{Electrophoretic mobility of ions and bulk conductivity}\label{sec:model}

   \begin{figure}[h]
   	\begin{center}
   		\includegraphics[width=0.95\columnwidth]{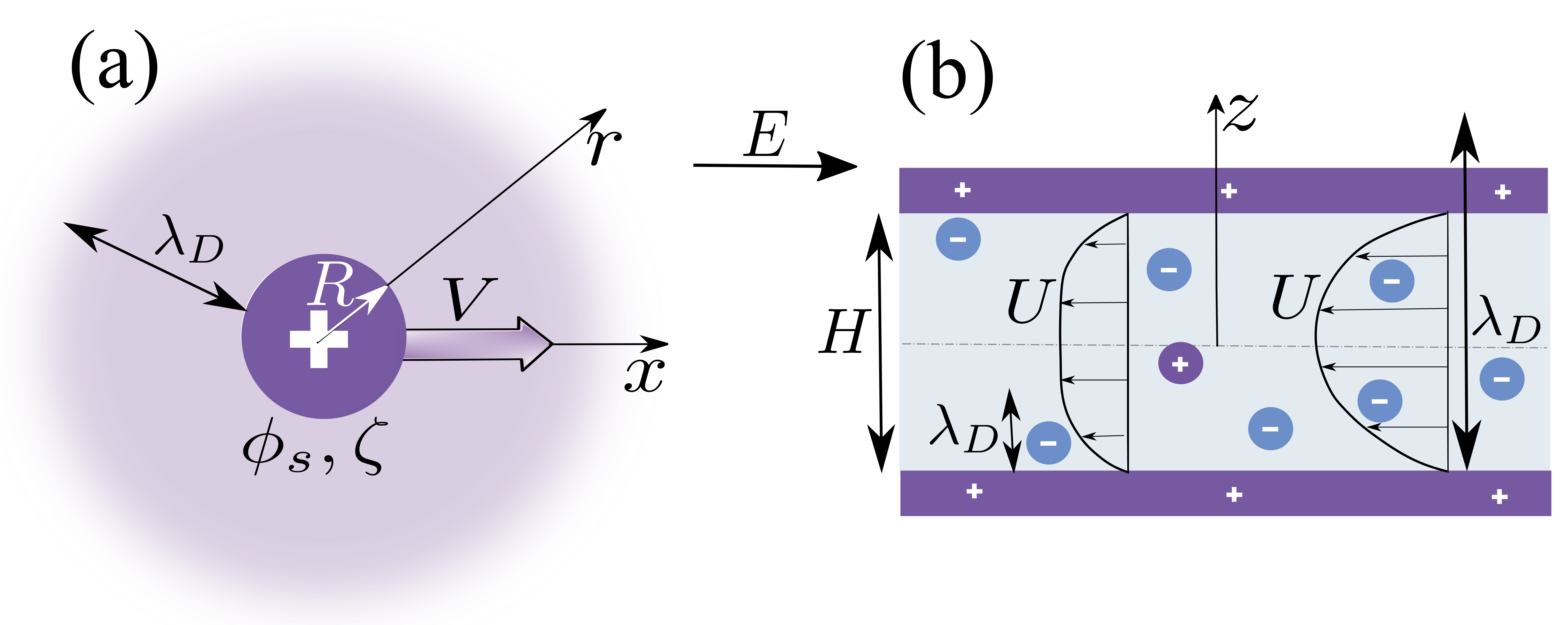}\\
   	\end{center}
   	\caption{(a) Sketch of the cation of hydrodynamic radius $R$ and surface potential $\phi_s$ in the bulk electrolyte solution characterized by  $\lambda_D$. The cation propels with the speed $V$ in the direction of the electric field $E$ and its mobility reflects the zeta potential $\zeta \neq \phi_s$ (b) Profiles of electro-osmotic  velocity $U$ induced in the positively charged channels of thickness $H$ subject to $E$ in the thick (left) and thin (right) channel modes.  }
   	\label{fig:sketch}
   \end{figure}

We consider a 1:1 salt solution of number density (concentration) $n_{\infty}$, dynamic viscosity $\eta $
and permittivity $\varepsilon $. Ions obey a Boltzmann distribution, $n_{\pm }(z)=n_{\infty}\exp (\mp \phi (z))$, where $\phi (z)=e\Phi(z)/(k_{B}T)$ is the dimensionless electrostatic
potential, $e$ is the elementary positive charge, $k_{B}$ is the Boltzmann
constant, $T$ is a temperature of the system, and the upper (lower) sign
corresponds to the positive (negative) ions. Note that CGS (Gaussian) electrostatic units are used throughout our paper. However, by using dimensionless energy units and expressing all characteristic lengths in terms of the Debye and Bjerrum lengths we will obtain the results that are independent of any specific system of units.
The Debye
screening length of an electrolyte solution, $\lambda_{D}=\left( 8\pi \ell _{B}n_{\infty}\right) ^{-1/2}$, is defined as usually with the Bjerrum
length, $\ell _{B}=\dfrac{e^{2}}{\varepsilon k_{B}T}$. Note that $\ell _{B}$ of water is equal to about $0.7$ nm for room temperature.  By analysing the experimental data it is more convenient to use the concentration $c_{\infty}[\rm{mol/l}]$, which is related to $n_{\infty} [\rm{m^{-3}}] $
as $n_{\infty} = N_A \times 10^3 \times c_{\infty}$, where $N_A$ is Avogadro's number.
We recall that a useful formula for 1:1 electrolyte is~\cite{israelachvili.jn:2011}
\begin{equation}\label{eq:DLength}
  \lambda_D [\rm{nm}] = \frac{0.305 [\rm{nm}]}{\sqrt{c_{\infty}[\rm{mol/l}]} },
\end{equation}
so that upon increasing $c_{\infty}$ from $10^{-5}$ to $1$ mol/l the screening length is reduced from about 100 down to ca. 0.3 nm.

For a description of electrostatic problems we will use a continuum mean-field (Poisson-Boltzmann) theory, which implies that ions are taken as point-like objects. As any approximation, the Poisson-Boltzmann formalism has its limits of validity, but it always describes very accurately the ionic distributions for monovalent ions up to $10^{-1}$ mol/l as long as the surfaces are not too highly charged~\cite{poon.w:2006}. However, since some of our results could have validity beyond this particular range of concentrations, we here expend the concentration range up to a few mol/l. While the mean-field approach could give a rather crude description  of potentials  in so concentrated bulk solutions by neglecting correlations and finite ion size it will provide us with an insight into the various factors that determine electrostatic functions.

By contrast, to describe the electrophoretic mobility of ions, it is always necessary to introduce their hydrodynamic radius into the theory. Inorganic ions have a hydrodynamic radius from 0.1 to 0.3 nm, except for a hydroxyl and a hydrogen ($R=0.047$ nm and $0.027$ nm, respectively)~\cite{kadhim.mj:2020}. Thus, the migrating ion is modeled here as a tiny spherical particle of unit positive or negative charge as sketched in Fig.~\ref{fig:sketch}(a).
When an  electric field $E$ is applied in the $x$-direction, an electro-osmotic flow around such spheres of unit charge is induced. The electroosmosis takes its  origin in the  EDL, where a tangential electric field generates a force that sets the fluid in motion. The emergence of this flow in turn provides hydrodynamic
stresses that cause the electrophoretic propulsion of the ions with a velocity $V_{\pm} = m_{\pm} E$. For simplicity, we assume that both types of ions have the same hydrodynamic radius $R$. In this case it is enough to consider only the case of cations of $m_+ = m$ (and correspondingly, $V_+ = V$). The mobility of anions will be the same, but taken with the opposite sign, $m_- = -m$.

The expression for $m$ can be written as
\begin{equation}\label{eq:M}
  m = \dfrac{\varepsilon Z}{4 \pi \eta } =  \dfrac{e }{4\pi \eta 	\ell _{B}}\zeta,
\end{equation}
where $Z$ is the electrokinetic or zeta potential of a spherical ion, which appears in \eqref{eq:M} via the Stokes equation, and $\zeta = e Z/k_B T$ is the dimensionless zeta potential. Note that Eq.\eqref{eq:M} implies that the ion propels in the direction of  $E$, if $\zeta$ is positive.

A knowledge of the zeta potential is thus sufficient to  determine the cation mobility. It has been well recognized in the last two decades that $\zeta$ depends on hydrodynamic boundary conditions at the surface~\cite{joly.l:2006,maduar.sr:2015,vinogradova.oi:2022}. For a hydrophilic flat wall the classical no-slip boundary condition holds, leading to $\zeta = \phi_s$ (see recent review~\cite{vinogradova.oi:2023} and references therein). For hydrophilic (spherical) ions  one can construct the solution for zeta potential in the form:
\begin{equation}\label{eq:zeta_general}
  \zeta = \mathcal{F} \phi_s,
\end{equation}
where $\mathcal{F}$ is the function of $\varrho = R/\lambda_D$. We return to $\mathcal{F}$ later by focussing first on calculations of $\phi_s$.

One can attribute to finite size ion the surface charge density $e/(4 \pi R^2)$, which is constant, i.e. independent on the salt concentration. The surface potential $\phi_s$, however, is established self-consistently and should generally depend on salt. The potential around such a sphere ($r \geq R$) satisfies the nonlinear Poisson-Boltzmann equation (NLPB), which in spherical coordinates reads
\begin{equation}
\dfrac{1}{r^{2}}\dfrac{d}{dr}\left( r^{2}\dfrac{d\phi }{dr}\right) =\lambda
_{D}^{-2}\sinh \phi.   \label{eq:NLPB}
\end{equation}
To integrate Eq.(\ref{eq:NLPB}) we impose two electrostatic boundary
conditions. The first condition requires that $\phi$ decays to zero as $r \to \infty$. The second condition is applied at the sphere surface
\begin{equation}
\phi^{\prime} |_{r=R}=-\dfrac{\ell _{B}}{R^2}. \label{eq:bc_CCs}
\end{equation}

The solution to Eq.\eqref{eq:NLPB} can be
obtained only numerically, but some approximate analytical results can been obtained the following two special cases:

\begin{itemize}
  \item The first is of low (volume) charge density case,  $\varrho^{2}\sinh \phi _{s} \ll 1$, where \eqref{eq:NLPB} reduces to the Laplace equation
  \begin{equation}  \label{eq:L_sphere}
\dfrac{1}{r^2} \dfrac{d}{d r} \left( r^2 \dfrac{d \phi}{d r} \right) \simeq 0.
\end{equation}
Using the boundary condition \eqref{eq:bc_CCs} the Laplace equation can be integrated analytically, yielding the following surface potential $\phi_s$ and outer potential profile:
\begin{equation}\label{eq:Laplace}
\phi_s \simeq \dfrac{\ell_{B}}{R}, \, \phi \simeq \dfrac{\ell_{B}}{r}.
\end{equation}
Thus, one can say that in this case $\phi_s = O (1)$ for inorganic ions, and $O(10)$ for hydroxyl and hydrogen. We thus exclude hydroxyl and hydrogen from the consideration since the surface potential appears  to be too high to apply the standard NLPB approach~\cite{borukhov.i:1997}.
We also conclude that
to a first-order approximation, for extremely small $\varrho$ the Coulombic interaction ($\propto r^{-1}$) is not screened or screened only slightly, even if $\phi_s$ is large.
  \item The second case is of the low potential,
  $\phi_s \leq 1$ (or $\Phi_s \leq 25$ mV), where \eqref{eq:NLPB} can be linearized resulting in the
Debye-H\"{u}ckel (DH) equation
\begin{equation}  \label{eq:DH_sphere}
\dfrac{1}{r^2} \dfrac{d}{d r} \left( r^2 \dfrac{d \phi}{d r} \right) \simeq \lambda _{D}^{-2} \phi.
\end{equation}
Performing the integration of \eqref{eq:DH_sphere} by imposing \eqref{eq:bc_CCs} then yields
\begin{equation}\label{eq:DH_solution}
 \phi_s \simeq \dfrac{ \ell_{B} }{ R \left(1+\varrho\right)}, \, \phi \simeq \dfrac{ \ell_{B} }{ \left(1+\varrho\right)} \dfrac{\mathrm{e}^{\varrho - r/\lambda_D}}{r}.
\end{equation}
When $\varrho$ is small, that is in dilute solutions, we obtain again $\phi_s = O (1)$. Strictly speaking, this should be beyond the validity of Eq.\eqref{eq:DH_sphere}. However, below we shall see that small $\varrho$ has repercussions for the range of applicability of Eq.\eqref{eq:DH_sphere}, which becomes no longer specific just to low potentials. Note that in contrast to the above case the potential ($\propto r^{-1} \mathrm{e}^{- r/\lambda_D}$) as a rule is strongly (exponentially) screened at distances larger than $\lambda_D$.
\end{itemize}

\begin{figure}[h]
\begin{center}
\includegraphics[width=0.9\columnwidth]{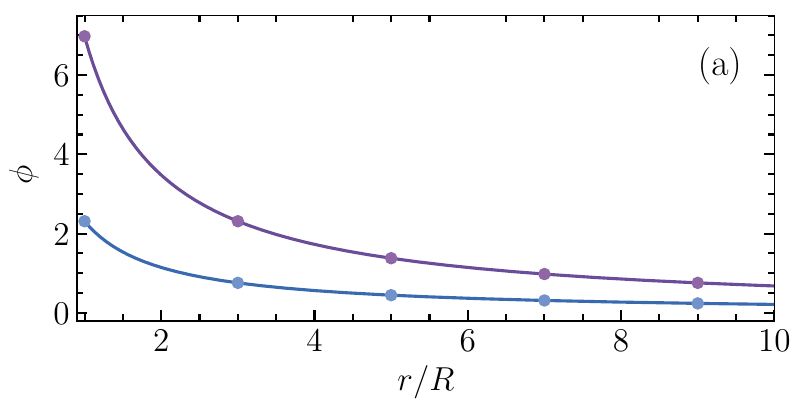}\\
\includegraphics[width=0.9\columnwidth]{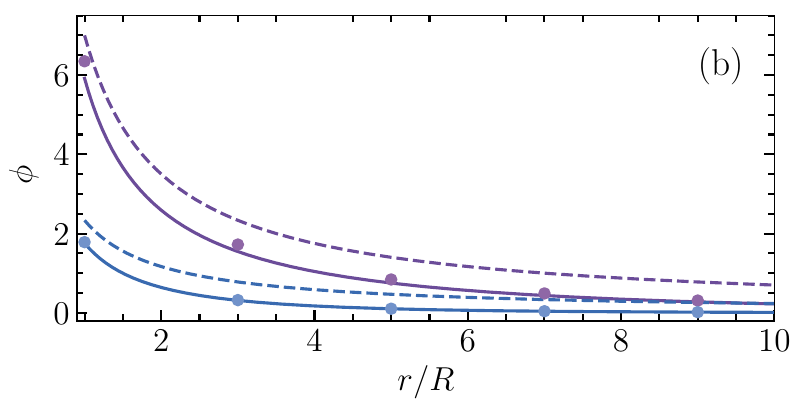}
\end{center}
\caption{Electrostatic potential profiles for cations of radius $R=0.1$ and $0.3$ nm (solid curves from top to bottom) calculated numerically using fixed $c_{\infty} = 10^{-4}$ mol/l (a) and $10^{-1}$ mol/l (b). Dashed curves indicate calculations from Eq.\eqref{eq:Laplace}. Calculations from Eq.\eqref{eq:DH_solution} are shown by circles. }
\label{fig:pot}
\end{figure}

Figure~\ref{fig:pot} shows the $\phi$-profiles for cations computed using  $R = 0.1$ and 0.3 nm. The first series of numerical calculations (Fig.~\ref{fig:pot}(a)) is made using $c_{\infty} = 10^{-4}$ mol/l that gives $\lambda_D \simeq 30$ nm, which corresponds to extremely small $\varrho$ for both ion sizes. Also included are calculations from Eqs.\eqref{eq:Laplace} and \eqref{eq:DH_solution}. It can be seen that they coincide with the numerical curves and with each other. The potential thus is not screened but is perfectly described by the DH theory.
In the second series (Fig.~\ref{fig:pot}(b)) we fixed $c_{\infty} = 10^{-1}$ mol/l or, equivalently, $\lambda_D \simeq 1$ nm. This results in $\varrho \simeq 0.1$ and $0.3$. The solution \eqref{eq:Laplace} to the Laplace equation now overestimates the potential, especially for smaller $R$, but Eq.\eqref{eq:DH_solution} provides quite good fits to the numerical data down to $r/R \simeq 3$. At smaller $r/R$ the predicted within the DH theory potential is  higher than computed, but only  quite slightly.
\begin{figure}[h]
\begin{center}
\includegraphics[width=0.9\columnwidth]{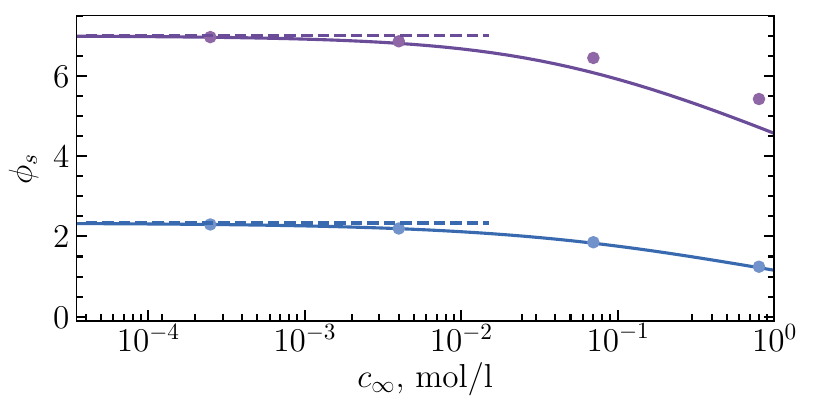}
\end{center}
\caption{Surface potential of the same cations as in Fig.~\ref{fig:pot} plotted as the function of $c_{\infty}$. Solid curves show the numerical results, dashed lines are obtained from Eq.~\eqref{eq:Laplace}, circles - from Eq.~\eqref{eq:DH_solution}. }
\label{fig:surf_pot}
\end{figure}

The curves for $\phi_s$ of the same cations  are plotted in Fig.~\ref{fig:surf_pot} as a function of $c_{\infty}$. For sufficiently dilute solutions $\phi_s$ remains constant and is well described by \eqref{eq:Laplace}, but at higher salt it decreases as $c_{\infty}$ grows. The numerical
data obtained at $R = 0.3$ nm are
well fitted by Eq.\eqref{eq:DH_solution} in all concentration range, but for ions of $R=0.1$ nm the fits are
quite good only for $c_{\infty} \leq 10^{-2}$ mol/l. In more concentrated solutions the predicted within the DH theory surface potential becomes less accurate and is larger than  computed. In other words, it is screened stronger than would be expected in the ``low potential'' case.   Nevertheless, the deviations of calculations from  Eq.\eqref{eq:DH_solution} from exact numerical results are much smaller than those made using Eq.\eqref{eq:Laplace}.

\begin{figure}[h]
\begin{center}
\includegraphics[width=0.9\columnwidth]{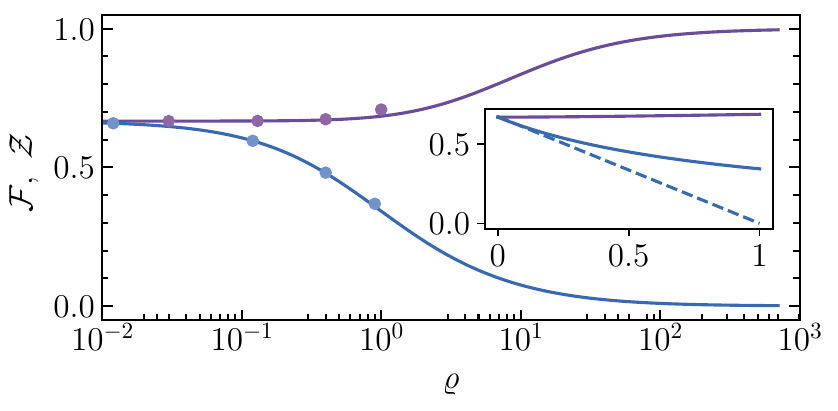}
\end{center}
\caption{Functions $\mathcal{F}$ (upper curve) and $\mathcal{Z}$ (lower curve) vs $\varrho$ calculated from Eqs.\eqref{eq:henry} and \eqref{eq:Z}. Circles show the same functions, but calculated using Eq.\eqref{eq:henry_lin}. Inset reproduces the region of $\varrho \leq 1$ in a lin scale. Dashed line corresponds to calculations from  Eq.\eqref{eq:Z_lin}.}
\label{fig:henry}
\end{figure}

We now turn to the function $\mathcal{F}$. In the general case of an arbitrary $\varrho$ and $\phi_s$ its analytical  calculation besets with difficulties and the numerical approaches have mostly been followed~\cite{obrien.rw:1978}. A simple expression, valid for any $\rho$, but imposing restrictions on a potential, has been derived by Henry~\cite{henry.dc:1931}
\begin{equation}\label{eq:henry}
  \mathcal{F} = 1 - e^{\varrho} \left[5 \Ei_7(\varrho) - 2 \Ei_5(\varrho)\right],
\end{equation}
where $\Ei_p (\varrho)= \varrho^{p-1} \Gamma (1-p,\varrho) = \varrho^{p-1} \int_{\rho}^{\infty} \frac{e^{-t}}{t^p} dt$ is the generalized exponential integral. A plot of $\mathcal{F}$ as a function of $\varrho$ is given in Fig.~\ref{fig:henry}. In the  (Smoluchowski) limit of $\rho \to \infty$ the second term in \eqref{eq:henry} vanishes and $\mathcal{F}(\varrho \to \infty) \to 1$. This situation is relevant to large colloid particles, but not small ions, where $\varrho \leq 1$. On reducing $\mathcal{\varrho}$ the value of $\mathcal{F}$ decreases. Since $\Ei_p (0) = 1/(p-1)$, it is easy to show that $\mathcal{F}(\varrho \to 0) \to 2/3$. This result is known as the H$\rm{\ddot{u}}$ckel limit~\cite{huckel.e:1924}. We stress that $\mathcal{F}$ is a special function, which cannot generally be reexpressed through elementary functions, but by Taylor expanding of \eqref{eq:henry} about zero of $\varrho$ one can derive
\begin{equation}\label{eq:henry_lin}
  \mathcal{F} \simeq \dfrac{2}{3} \left[1 + \left(\dfrac{\varrho}{4}\right)^2 \right].
\end{equation}

The calculations from \eqref{eq:henry_lin} are included in Fig.~\ref{fig:henry}. We see that the range of validity of this asymptotic result is amazingly  large, and Eq.\eqref{eq:henry_lin} can, in principle, be used up to $\varrho \simeq 1$.

Should the Henry equation be applied to calculate $\zeta$ of ions? This is equivalent to asking: is the use of \eqref{eq:henry}  justified enough with our magnitudes of ion potentials? The answer to this question is by no means obvious. Indeed, the Henry equation has been obtained using the DH theory and traditionally applies for $\phi_s \leq 1$ only. However, in essence, Eq.\eqref{eq:henry} is derived
from the general integral expression for mobility by substituting  an exponentially decaying  potential \eqref{eq:DH_solution}~\cite{ohshima.h:1983}. As demonstrated above, when $\varrho \leq 1$, it is not necessary to make assumptions about low surface potential to provide a sufficient accuracy of Eq.\eqref{eq:DH_solution}.
Thus we might argue that \eqref{eq:henry} should be a sensible approximation for small ions although their surface potential is not low. From Eqs.\eqref{eq:zeta_general} and \eqref{eq:DH_solution} it follows then that
 the zeta potential of an univalent cation is given by
\begin{equation}\label{eq:zeta_ion}
  \zeta \simeq  \dfrac{\ell_{B}}{R} \mathcal{Z},
\end{equation}
where
\begin{equation}\label{eq:Z}
  \mathcal{Z} \simeq  \dfrac{\mathcal{F}}{1+\varrho}.
\end{equation}

The upper possible values of the zeta potential is attained when $\rho \to 0$ (infinite dilution), where the H$\rm{\ddot{u}}$ckel limit, $\mathcal{Z} \to 2/3$,
is recovered
\begin{equation}\label{eq:zeta_ion_max}
  \zeta \to  \zeta_0 \simeq \dfrac{2\ell_{B}}{3R}.
\end{equation}

For small $\varrho$ Eq.\eqref{eq:Z}  can be expanded about $\varrho = 0$ and, to first order in $\rho$
\begin{equation}\label{eq:Z_lin}
  \mathcal{Z} \simeq \dfrac{2}{3} (1 - \varrho ).
\end{equation}
This implies that if $\varrho \ll 1$ (very high dilution), the decrease in $\mathcal{Z}$ is caused only by the reduction of the surface potential with salt concentration, but not by the variation in $\mathcal{F}$, which (to leading order) remains constant.

Finally, the simplest expression that should be an excellent approximation at $\varrho \leq 1$, can be obtained by substituting Eq.\eqref{eq:henry_lin} to \eqref{eq:Z}
\begin{equation}\label{eq:henry_lin2}
  \mathcal{Z} \simeq \dfrac{2}{3} \left[ \dfrac{1 + \left(\dfrac{\varrho}{4}\right)^2}{1+\varrho} \right].
\end{equation}
By analogy to prior definitions one can introduce the term \emph{low dilution} specifying the situation $0.1 \leq \varrho \leq 1$.

The  $\mathcal{Z}$-function calculated from \eqref{eq:Z} using $\mathcal{F}$ given by Eq.\eqref{eq:henry} is included in Fig.~\ref{fig:henry}. It can be seen that it decays from $\mathcal{Z}(\rho \to 0) \to 2/3$ down to zero as $\varrho \to \infty$. Also included are results for $\mathcal{Z}$ obtained using $\mathcal{F}$ calculated from \eqref{eq:henry_lin}, which confirm that using approximate $\mathcal{F}$  provides excellent agreement with exact $\mathcal{Z}$ up to $\varrho \simeq 1$. To examine the region of $\rho \leq 1$  more closely, the  $\mathcal{F}$ and $\mathcal{Z}$ curves are reproduced in the inset
 along with $\mathcal{Z}$ calculated from Eq.\eqref{eq:Z_lin}. The inset is intended to illustrate the
range of validity of this asymptotic result. It is well seen that Eq.\eqref{eq:Z_lin} is applicable only if $\varrho \ll 1$. For larger $\varrho$ calculations from \eqref{eq:Z_lin} are grossly inaccurate and differ significantly from predictions of Eq.\eqref{eq:Z}, by underestimating the value of $\mathcal{Z}$. At $\varrho \geq 1$ they would even  lead to zero or negative $\mathcal{Z}$. This automatically restricts the use of \eqref{eq:Z_lin} to very small concentrations. We return to this defect in the linear theory at a later stage to show that in some situations it can be partially offset, but in others leads to misleading conclusions.

\begin{figure}[h]
\begin{center}
\includegraphics[width=0.9\columnwidth]{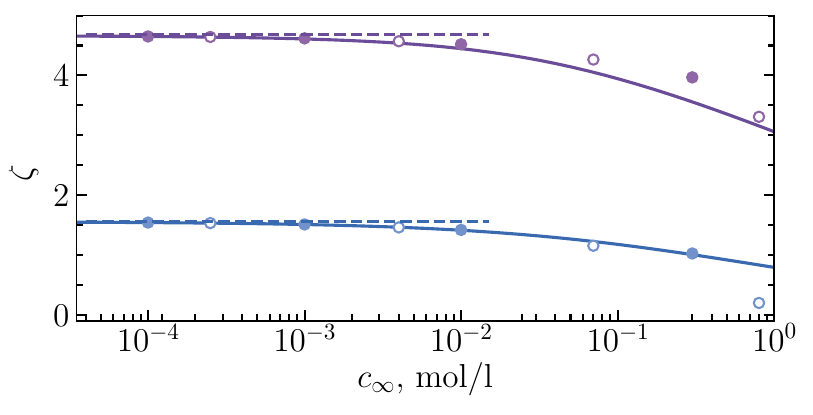}
\end{center}
\caption{Zeta potentials calculated from Eq.~\eqref{eq:zeta_general} for the same parameters as in Fig~\ref{fig:surf_pot}. Dashed lines are obtained using Eq.~\eqref{eq:zeta_ion_max}. Filled and open circles correspond to Eq.~\eqref{eq:zeta_ion} with $\mathcal{Z}$ calculated from Eqs.~\eqref{eq:Z} and~\eqref{eq:Z_lin}. }
\label{fig:zeta}
\end{figure}

The zeta potential can thus be found from Eq.\eqref{eq:zeta_general}  using computed (exact) $\phi_s$ and $\mathcal{F}$ defined by Eq.\eqref{eq:henry}. For analytical calculations Eq.\eqref{eq:zeta_ion} could be employed with $\mathcal{Z}$ given by \eqref{eq:Z}. Clearly, $\mathcal{Z}$ can be calculated with Eq.\eqref{eq:Z_lin} provided $\varrho \ll 1$ or, if $\varrho \leq 1$, from \eqref{eq:henry_lin2}. Figure~\ref{fig:zeta} includes zeta potentials computed for the same values of $R$ as before  along with the straight lines corresponding to $\zeta_0$ given by \eqref{eq:zeta_ion_max}. The latters describe perfectly the distinct plateau regions observed at low salt concentrations. On reducing $c_{\infty}$ further $\zeta$ decreases, and we see that the use of \eqref{eq:henry_lin2} provides an excellent fit to the $\zeta$-curve for cations of $R = 0.3$ nm. However, \eqref{eq:Z_lin} significantly underestimates the value of $\zeta$ at high concentrations (above $10^{-1}$ mol/l, where $\varrho \geq 0.3$), as expected. For ions of $R = 0.1$ nm the value of $\varrho$ at a given $c_{\infty}$ is three times smaller, so the approximations should be applicable in the larger range of  concentrations. We see, however, that the data
obtained using \eqref{eq:Z_lin} and  \eqref{eq:henry_lin2} show approximate  $\zeta$ that is higher than numerical when $c_{\infty} \geq 10^{-1}$ mol/l. An explanation for this discrepancy is obviously an overestimated surface potential shown in Fig.~\ref{fig:surf_pot}. Simultaneously, Eq.\eqref{eq:Z_lin} underestimates $\mathcal{Z}$, so these two discrepancies of first-order calculations partly compensate each other. As a result, in concentrated solutions Eq.\eqref{eq:Z_lin} even appears to be slightly more accurate than \eqref{eq:henry_lin2}.

Substituting \eqref{eq:zeta_ion} into \eqref{eq:M}, we derive for an electrophoretic mobility of cations
\begin{equation}\label{eq:M_ion_full}
  m \simeq \dfrac{e }{4\pi \eta R} \mathcal{Z}.
\end{equation}

The  ion mobility attains its upper value when $\rho \to 0$, where $\mathcal{Z} \to 2/3$
and $m \to m_0$ given by Eq.\eqref{eq:inf}. Thus  we have recovered $\zeta$ in the H$\rm{\ddot{u}}$ckel limit and justified the assumption of the Stokes drag, although this derivation  differs  from  the  conventional  arguments. Our treatment  clarifies  the status of Eqs.\eqref{eq:zeta_ion_max} and Eq.\eqref{eq:inf}. They constitute a rigorous asymptotic results for zeta potential and electrophoretic mobility of ions in the limit $\rho \to 0$, which  applies provided the electrolyte solution is sufficiently dilute for a potential to be described by the Laplace equation, Eq.\eqref{eq:L_sphere}. In other words, a so-called infinite dilution is nothing more than the (quite large) range of concentrations, where the application of the Laplace equation is justified.

A bulk conductivity is given by
\begin{equation}\label{eq:Kfull}
  K_{\infty} \simeq  \dfrac{e^2 n_{\infty}}{2 \pi \eta R}  \mathcal{Z} \simeq \dfrac{3}{2} K_{\infty}^0 \mathcal{Z}.
\end{equation}
Clearly, the last equation reduces to \eqref{eq:Kinf} only in the limit $\varrho \to 0$.
Consequently, for small $\varrho$, when $\mathcal{Z}$ is expressed by \eqref{eq:Z_lin}, Eq.\eqref{eq:Kfull} reduces to
\begin{equation}\label{eq:Klin}
  K_{\infty} \simeq K_{\infty}^0 \left(1 - \varrho\right) \simeq K_{\infty}^0 - \dfrac{2 e^2  \left( 2\pi \ell _{B} \right)^{1/2}}{3 \pi \eta }   n_{\infty }^{3/2}.
\end{equation}
Note that the second term $ \propto - n_{\infty }^{3/2}$ and does not depend on $R$. The form of \eqref{eq:Klin} is identical to Eq.\eqref{eq:Kolr}, but the  constant $\alpha$ is interpreted. Although Kohlrausch's equation has been interpreted in similar fashion before, its nature is apparent now. This equation is a consequence of  a linearization of the ion zeta potential at small $\varrho$ and, hence, takes into account solely a decrease in $\phi_s$ with salt, by neglecting an increase in $\mathcal{F}$. Finally, we note that the bulk conductivity given by \eqref{eq:Klin} takes its maximum value at $n_{\infty }\simeq (18 \pi \ell _{B} R^2)^{-1}$. For ions of $R=0.3$ nm this corresponds to $c_{\infty }\simeq 0.5$ mol/l (and $R=0.1$ nm would give  $c_{\infty }\simeq 4.2$ mol/l). Thus it is tempting to speculate that the maxima in conductivity observed in
experiments and simulations at molar salt concentrations~\cite{fong.kd:2020,zhang.w:2020} do arise naturally from the electrophoretic model we discuss.
In this context it is important to recall that the NLPB description and restrictions on $\varrho$ that led to Eq.\eqref{eq:Klin}, are only appropriate when concentrations are much smaller than expected for this maximum to occur. Indeed, the calculation using Eq.\eqref{eq:Kfull} shows that  $K_{\infty}$  has no maximum, thus its appearance at high concentrations in the Kohlrausch's model, which is restricted by very high dilution, is just an artefact of the linearization of $\mathcal{Z}$. 

If we make the assumption that $\varrho \leq 1$, Eq.~\eqref{eq:Kfull} reduces to
\begin{equation}\label{eq:Knew}
  K_{\infty} \simeq K_{\infty}^0 \left[ \dfrac{1 + \left(\dfrac{\varrho}{4}\right)^2}{1+\varrho} \right] \simeq K_{\infty}^0  \left[ \dfrac{1 +  \dfrac{R^2 \pi \ell _{B} n_{\infty }}{2}}{1 + R \left( 8\pi \ell _{B} n_{\infty }\right)^{1/2}} \right]
\end{equation}
This  equation that has been rigorously derived here should provide a more accurate estimation of the bulk conductivity than known theories and a practically exact description of $K_{\infty}$ for solutions composed of largest inorganic ions (of $R=0.3$ nm).

\begin{figure}[h]
\begin{center}
\includegraphics[width=0.9\columnwidth]{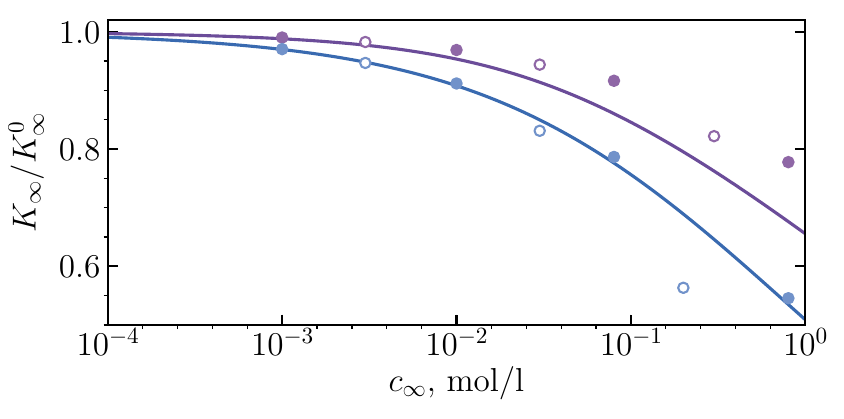}
\end{center}
\caption{Bulk conductivity vs $c_{\infty}$ computed for $R=0.1$ nm (top set of curves) and $R=0.3$ nm (bottom set of curves). $K^{0}_{\infty}$ is shown by the dashed curve. Filled and open circles indicate calculations from Eqs.~\eqref{eq:Kfull} and~\eqref{eq:Klin}.}
\label{fig:K_bulk}
\end{figure}

Figure~\ref{fig:K_bulk} shows $K_{\infty}/K_{\infty}^0$ as a function of $c_{\infty}$. As follows from \eqref{eq:Kfull} this ratio is equal to $3 \mathcal{Z}/2$, so it is enough to calculate $\mathcal{Z}$. The calculations from Eq.\eqref{eq:Z} are presented and compared with those from \eqref{eq:Z_lin} and \eqref{eq:henry_lin2}. An overall conclusion from this plot is that the bulk conductivity reduces with salt and is generally smaller than predicted by Eq.\eqref{eq:Kinf}. The deviations from $K_{\infty}^0$ increase with the hydrodynamic radius of ions. For univalent ions of $R=0.3$ nm, Eq.\eqref{eq:Klin} is justified provided $\lambda_D \geq 3$ nm ($c_{\infty} \leq 10^{-2}$ mol/l). In more concentrated solutions, Kohlrausch's law should underestimate the bulk conductivity, and this is indeed well seen in Fig.~\ref{fig:K_bulk}. However, the theoretical data obtained from Eq.\eqref{eq:henry_lin2}, which is valid when $\lambda_D \geq 0.3$ nm ($c_{\infty} \leq 1$ mol/l), practically coincide with the exact results. For ions of $R=0.1$ nm, the ranges of validity of Eqs.\eqref{eq:Z_lin} and \eqref{eq:henry_lin2} are larger, but we see that at $c_{\infty} \geq 10^{-2}$ mol/l there is some discrepancy from Eq.\eqref{eq:Z}. The discrepancy is always in the direction of the smaller reduction in conductivity than predicted by \eqref{eq:Z}, and is practically the same for both approximations. As discussed above, at high concentrations the Henry electrostatic arguments fail a little  for this ion size since $\phi_s$ obtained within the DH theory slightly exceeds the exact one. Of course this immediately raises an issue about the accuracy of the simple Henry expression \eqref{eq:henry}  that we have used here, so for ions of $R = 0.1$ nm in concentrated solutions it can only be considered as a first-order approximation.
Nevertheless, the calculation presented here demonstrates the effect of salt and the magnitude of the reduction in conductivity. Surprisingly, calculations also justify the use of Kohlrausch's equation for inorganic ions of the smallest hydrodynamic radius, although it fails for large ions of $R = 0.3$ nm. It must be remembered that
this will be a first-order calculation only, and we do not expect it to be very accurate.

In Appendix~\ref{a:1} we compare our results with some experimental data (for NaCl and LiClO$_4$) in the large concentration range (up to 3 mol/l). An overall conclusion is that the agreement with experiment is extremely good. Moreover, at molar concentrations our simple theory provides a better fit to the conductivity data than recently suggested model~\cite{avni.y:2022}.

\section{Conductivity of  electrolyte in channels}\label{sec:channel}

One imagines now a symmetric slit of thickness $H$ in contact with  bulk reservoir of a 1:1 salt solution of concentration $c_{\infty}$ subject to an
electric field $E$ in the $x$ direction as sketched in Fig.~\ref{fig:sketch}(b). The axis $z$ is defined normal to the surfaces of dimensionless potential $\Psi_{s}$
and charge density $\sigma $ located at $z= \pm H/2$. Without loss of generality, the surface charges are taken as cations ($\sigma$ is positive).
For a symmetric planar channel of thickness $H$, it is enough to consider $z = H/2$ because
of the $z \leftrightarrow - z$ symmetry. Our discussion  applies for channels down to a few nanometers to ignore various effects that would be important for thinner channels (see~\cite{vinogradova.oi:2023} and references therein).

The electrostatic potential in the slit satisfies the nonlinear Poisson-Boltzmann
equation. In the Cartesian coordinates
\begin{equation}  \label{eq:NLPB2}
\psi^{\prime \prime} =\lambda _{D}^{-2}\sinh \psi,
\end{equation}
where $^{\prime}$ denotes $d/d z$.

To integrate Eq.(\ref{eq:NLPB2}) we impose two electrostatic boundary
conditions. The first condition always reflects the symmetry of the channel $%
\psi^{\prime} |_{z=0}=0$. The second condition is applied at the walls
and can be either that of a constant surface potential (conductors)
\begin{equation}
\psi |_{z=H/2}=\psi _{s},  \label{eq:bc_CP}
\end{equation}
or of a constant surface charge density (insulators)
\begin{equation}
\psi^{\prime} |_{z=H/2}=\dfrac{2}{\ell _{GC}},  \label{eq:bc_CC}
\end{equation}
where $\ell _{GC}=\dfrac{e}{2\pi \sigma \ell _{B}}$ is the Gouy-Chapman
length. These situations are referred below to as CP and CC cases.

A tangential electric field $E$ induces an electro-osmotic flow of velocity $U(z)$ that satisfies Stokes' equations with an electrostatic body force:
\begin{equation}  \label{eq:Stokes}
  u^{\prime \prime} = \psi^{\prime \prime},
\end{equation}
where $u(z) = \dfrac{4\pi \eta
\ell _{B}}{e E} U(z)$ is the dimensionless fluid velocity. The fluid velocity at $z=H/2$ satisfies no slip boundary conditions $u|_{z=H/2}=0$. Performing the integration in Eq.\eqref{eq:Stokes} with prescribed boundary conditions yields
\begin{equation}
u(z)=\psi (z)-\psi _{s}.
\label{eq:Stokes_solution}
\end{equation}

An applied electric field also generates an electric current in the channel. In contrast to a bulk case, the local current density $j(z)$ is not uniform, so it is convenient to describe the system in terms of average values. We denote a mean current density as $J$ and a mean
channel conductivity as $K = J/E$. An average value of any other function $f$ will be denoted as%
\begin{equation}\label{eq:average}
\langle f \rangle = \frac{2}{H}\int\limits_{0}^{H/2}\,fdz.
\end{equation}

 To determine $K$ a knowledge of $J$ is required. Below we calculate $J$ assuming a weak field, so that in steady state $\psi(z)$ is independent of the fluid flow.
 The local current density of thermal ions in confined electrolyte is $j_{+}+j_{-}$, where $j_{\pm }=\pm en_{\pm }(U \pm m E)$. The first term is associated with the convective contribution, i.e. with the transport of ions by an electro-osmotic flow of velocity $u$ given by Eq.\eqref{eq:Stokes_solution}. The second term represents a migration contribution caused by the (electrophoretic) propulsion of ions with respect to the solvent.
The mean current density in the channel is then
\begin{equation}
  J = \langle j_{+} \rangle + \langle j_{-} \rangle,
\end{equation}
where
\begin{equation}
 \langle j_{\pm} \rangle = \pm\dfrac{2e E n_{\infty}}{H} \int_{0}^{H/2} e^{\mp\psi} \left[ m_e \pm m  \right] dz,
\end{equation}
and $m_e$ is the electro-osmotic mobility of the channel~\cite{vinogradova.oi:2022}
\begin{equation}\label{eq:Stokes_solution_EO_noslip2}
m_e (z) = - \dfrac{e}{4\pi \eta \ell _{B}} (\psi _{s} - \psi
 (z)).
\end{equation}%
This equation implies that in a positively charged channel the electro-osmotic flow is directed against $E$.

Substituting $m_e$ and making use of \eqref{eq:NLPB2} yields
\begin{equation}\label{eq:Jchannel1}
  J =  \dfrac{e^2 E n_{\infty}}{\pi \eta \ell _{B} H}  \zeta \int_{0}^{H/2} \left(\dfrac{\lambda_D^2}{\zeta} (\psi _{s} - \psi)  \psi'' + \cosh \psi \right) dz,
\end{equation}
where the first term in the integrand is associated with convective contribution and the second with the migration (electrophoresis) of ions. Performing the integration of the convective term by parts one can reduce \eqref{eq:Jchannel1} to
\begin{equation}\label{eq:Jchannel2}
  J =  \dfrac{e^2 E n_{\infty}}{2\pi \eta \ell _{B}}  \zeta \left[ \dfrac{ \lambda_D^2}{\zeta} \langle(\psi^{\prime})^2\rangle + \langle\cosh \psi\rangle \right],
\end{equation}
which comprises of the terms incorporating the mean electrostatic energy $\propto \langle(\psi^{\prime})^2\rangle$ and entropy $\propto \langle\cosh \psi\rangle$ of the ions in the channel.

First integration of Eq.\eqref{eq:NLPB2} from the mid-plane position to an arbitrary $z$ gives
 \begin{equation}\label{eq:coshm}
\langle\cosh \psi\rangle = \cosh \psi_m +\dfrac{\lambda_D^{2} \langle(\psi^{\prime})^2\rangle}{2},
\end{equation}
where $\psi_m$ is the potential at the midplane ($z=0$), where the electric field vanishes. This implies that
$ \langle(\psi^{\prime})^2\rangle$ affects both convective and migration contributions in \eqref{eq:Jchannel2}.

From Eqs.\eqref{eq:coshm} and Eq.\eqref{eq:Jchannel2} it follows that the expression for a mean current density
\begin{equation}
  J =  \dfrac{e^2 E n_{\infty}}{2\pi \eta \ell _{B}}  \zeta \left[ 1+ \Pi + \dfrac{\lambda_D^{2} \langle(\psi^{\prime})^2\rangle}{2}\left(1+\dfrac{2}{\zeta} \right) \right],
\end{equation}
where $\Pi = \cosh \psi_m - 1$ is an excess of dimensionless osmotic pressure at the mid-plane compared to the bulk solution, which represents an electrostatic disjoining pressure~\cite{derjaguin.bv:1941}. Note that only the second term in the expression in round brackets is of convective (electro-osmotic) origin, all other contributions are due to electrophoretic migration of ions.
The disjoining pressure is  expected to approach
its lower bound $\Pi=0$ when EDLs do not overlap ($H\to \infty$), and to the upper bound, $\Pi=\cosh \psi_s - 1$, provided the overlap is so strong ($H\to 0$) that the potential across the slit remains uniform and equal to $\psi_s$. Consequently, $\psi^{\prime} \to 0$, which implies that the convective contribution to $K$ vanishes. Thus the amplification of conductivity is caused only by the enhanced  electrophoretic migration. The latter is due to an enhanced concentration of ions in the channel compared to the bulk phase. Later we shall see that the conductivity in some situations can indeed be expressed simply through $\cosh \psi_s$.
 Note that some approximate expressions for a conductance corresponding to this upper bound on $\Pi$ have been derived~\cite{biesheuvel.pm:2016,peters.pb:2016} and that this case is often  referred in literature to as ``Donnan equilibrium''.

The expression for the conductivity then reads
\begin{equation}\label{eq:Kfinal}
  K =  \dfrac{e^2 n_{\infty}}{2 \pi \eta \ell _{B}}  \zeta \left[ 1+ \Pi+\dfrac{\lambda_D^{2} \langle(\psi^{\prime})^2\rangle}{2}\left(1+\dfrac{2}{\zeta} \right) \right].
\end{equation}
The first term is the bulk conductivity that augments with salt. The second and third terms is the additional conductivity due to EDLs.
This equation is general and valid for any $H$ as well as for any model of $\zeta$. In our model, $ \zeta$ and $K_{\infty}$ are given by Eq.\eqref{eq:zeta_ion} and
Eq.\eqref{eq:Kfull}. Thus, to determine $K$ from Eq.\eqref{eq:Kfinal} we only have to substitute the relevant expressions for $\langle(\psi^{\prime} )^{2}\rangle$ and disjoining pressure $\Pi$. In the general case they can be calculated only numerically, but one can obtain closed-form analytical expressions for the two experimentally relevant modes specified by a relation between $\ell _{GC}$ and $\psi_s$, known as the contact theorem:

\begin{itemize}
  \item In the so-called \emph{thick channel mode} the contact theorem coincides with the Grahame
equation for a single wall~\cite{vinogradova.oi:2021}
 \begin{equation}\label{eq:pot-charge_hs}
\psi _{s} = 2\arsinh\left(\frac{\lambda _{D}}{\ell _{GC}}\right).
\end{equation}
The ratio $\lambda_D/\ell _{GC} \propto \sigma n_{\infty}^{-1/2}$ reflects the \emph{effective surface charge}. The surfaces are referred to as \emph{weakly charged} when $\lambda_D/\ell _{GC} \leq 1$, and to as \emph{strongly charged} if $\lambda_D/\ell _{GC} \gg 1$. The Grahame
equation is valid not only for thick channels, $H/\lambda_D \gg 1$, but also for strongly charged channels of $H/\lambda_D = O(1)$ termed quasi-thick.
    \item In the \emph{thin channel mode} the contact theorem reads~\cite{silkina.ef:2019}
\begin{equation}
\psi _{s}\simeq  \arsinh \left(\dfrac{2 \ell_{Du}}{H}\right),
  \label{eq:pot-charge_ls}
\end{equation}
where
\begin{equation}  \label{eq:dukhin_length}
\ell_{Du} =  \frac{\lambda_{D}^2}{\ell_{GC}} = (8 \pi \ell_{GC} \ell_{B} n_{\infty})^{-1}
\end{equation}
is the Dukhin length, which in very dilute solutions can be larger than any conceivable Debye length.
 Clearly, the linearization of \eqref{eq:pot-charge_ls} cannot be justified  when $\ell_{Du}/H \geq 1$, even provided the effective charge is small.
 Although recent consideration supposed that this mode is restricted by weakly charged thin channels~\cite{vinogradova.oi:2021}, the derivation of Eq.\eqref{eq:pot-charge_ls} only imposes $\phi_s > H/(2 \ell _{GC})$. Thus, it is valid not only in the thin channel limit, $H/\lambda_D \ll 1$, but also for
 quasi-thin films of $H /\lambda_D = O(1)$ provided
the effective charge is below $2 \psi_s \lambda_D/H$ (but not necessarily low) or that $H/\lambda_D < 2 \ell _{GC} \psi_s/\lambda_D$.
\end{itemize}

\subsection{Thick channel mode}\label{sec:ThickCR}

The expression for $\langle(\psi^{\prime} )^{2}\rangle$ in the thick channel mode has been obtained before~\cite{vinogradova.oi:2021} and the contribution of the disjoining pressure can safely be neglected:
\begin{equation}\label{eq:average_dphi}
\langle(\psi^{\prime} )^{2}\rangle =   \dfrac{8}{H \lambda_D} \left(\sqrt{1+\dfrac{\ell _{Du}}{\ell _{GC}}} - 1\right), \Pi \simeq 0.
\end{equation}
When $\ell _{Du}/\ell _{GC}$ is large, Eq.\eqref{eq:average_dphi} reduces to
\begin{equation}
\langle(\psi^{\prime} )^{2} =   \dfrac{8}{H \ell _{GC}},
\end{equation}
whence
\begin{eqnarray}
K &\simeq& K_{\infty} \left[ 1 + \dfrac{4 \ell_{Du}}{H}\left(1+\dfrac{2}{\zeta} \right)\right]\label{eq:hydrophilic_thickCC}\\
&\simeq& K_{\infty} \left[ 1 +   \dfrac{\lambda_D}{H} \sinh \dfrac{\psi _{s}}{2}\left(1+\dfrac{2}{\zeta} \right)\right]\label{eq:hydrophilic_thickCP}
\end{eqnarray}

\begin{figure}[h]
	\begin{center}
		\includegraphics[width=0.9\columnwidth]{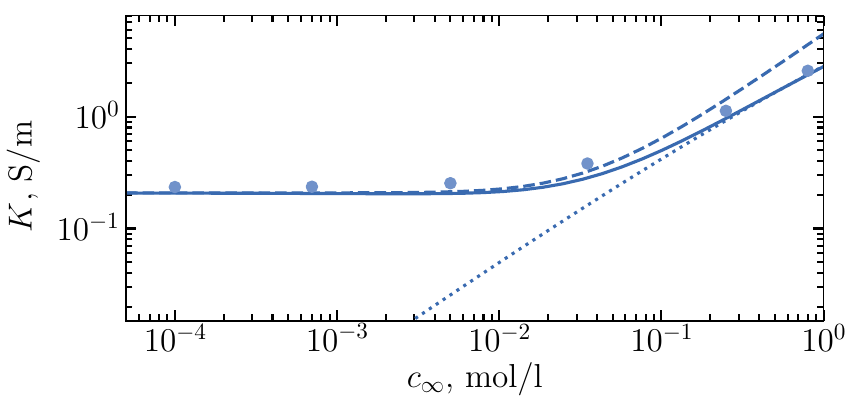}
	\end{center}
	\caption{Conductivity vs $c_{\infty}$ computed for a CC channel of $H=20$ nm and $\ell_{GC} = 1$ nm using $R=0.3$ nm. Solid curve shows calculations from Eq.~\eqref{eq:Kfinal}. Dashed curve represents $K$ calculated using ionic electrophoretic mobility from Eq.~\eqref{eq:inf}. Dotted curve corresponds bulk conductivity calculated from Eq.~\eqref{eq:Kfull}. Circles are obtained from Eqs.~\eqref{eq:hydrophilic_thickCC}.}
	\label{fig:K_cc}
\end{figure}

As an example we consider a CC channel. If for a channel of $H = 20$ nm we keep $\ell_{GC}$  fixed, on varying $c_{\infty}$ it is possible to obtain the conductivity curves displayed in Fig.~\ref{fig:K_cc}. For this calculation we use $\ell_{GC} = 1$ nm ($\sigma \simeq 36 $ mC/m$^2$).  With these parameters the thick channel mode  holds in the whole concentration range, even when the channel is formally thin, $H \ll \lambda_D$ (or $c_{\infty} \leq 10^{-4}$ mol/l). We now make no attempt to compare ions of different hydrodynamic radius by presenting the results only for $R=0.3$ nm. Instead, Fig.~\ref{fig:K_cc} is intended to illustrate the applicability of \eqref{eq:hydrophilic_thickCC} and to indicate the range of validity of approximations for $\zeta$ that can be  employed. In this context it is important to recall that for ions of radius 0.3 nm our theoretical description is appropriate in the very large concentration range and the derived
approximate expression for $\zeta$, Eq.\eqref{eq:zeta_ion}, becomes very accurate.
The computed $K$ has been obtained from  Eq.\eqref{eq:Kfinal} using $\psi$ obtained from the numerical solution of \eqref{eq:NLPB2}. The calculations are made using $\zeta$ calculated from Eq.\eqref{eq:zeta_general} with the exact $\phi_s$ and $\mathcal{F}$ given by \eqref{eq:henry} and $\mathcal{F}=2/3$. It can be seen that  at sufficiently small concentrations both curves coincide and $K$ remains constant that is independent on $c_{\infty}$. The plateau occurs when the EDL conductivity dominates over the bulk one, which happens at $\ell_{Du}/{H} \geq 1$, and at this branch of the conductivity curve $\zeta$ is described by \eqref{eq:zeta_ion_max} as seen in Fig.~\ref{fig:zeta}. Inserting \eqref{eq:zeta_ion_max} into Eq.\eqref{eq:hydrophilic_thickCC}, we reproduce the expression for a conductivity in thick channel regime obtained in the assumption of an infinite dilution (Stokes mobility of ions)~\cite{vinogradova.oi:2021}
\begin{equation}
  K \simeq K_{\infty}^0 \dfrac{4 \ell_{Du}}{H}\left(1+\dfrac{3 R}{\ell_B} \right)
\end{equation}
 The plateau thus appears since for a CC channel $K \propto K_{\infty }^0 \ell _{Du}$ does not depend on $c_{\infty}$, and the conductivity amplification at the plateau branch is huge since it scales as $\propto \ell_{Du}/H$. Say,  $c_{\infty} = 10^{-3}$ mol/l ($\lambda_D \simeq 10$ nm) gives for our channel $ \ell_{Du}/H \simeq 5$, so that at this concentration the conductivity is amplified in more than 30 times compared to a bulk one. We recall that the first term in the round brackets is due to a migration of ions, while the second is caused by their convective transfer. Since $3 R/\ell_B \simeq 1.3$ we conclude that these two contributions determining the hight of the conductivity plateau in Fig.~\ref{fig:K_cc} are of the same order of magnitude,  so neither of them can be ignored, but the convection effect is slightly more pronounced. This conclusion certainly refers only to ions of $R=0.3$ nm. If we would take ions of $R=0.1$ the convective contribution to the plateau height would be 0.4 of migrational.  For larger $c_{\infty}$ all $K$-curves increase with $c_{\infty}$ by becoming separated. The curve obtained using the Henry expression for $\mathcal{F}$ converges to the bulk conductivity  calculated from Eq.\eqref{eq:Kfull}. The theoretical results obtained from \eqref{eq:hydrophilic_thickCC} using Eq.\eqref{eq:zeta_ion}  are also included in Fig.~\ref{fig:K_cc}.   We conclude that they are in excellent agreement with numerical data for all $c_{\infty}$ including both the plateau and the ``bulk'' branch of the conductivity curve. Returning to numerical results, a similar curve obtained in the H$\rm{\ddot{u}}$ckel limit, $\mathcal{F} = 2/3$, increases faster and with different slope, by giving overestimated $K$ at high  concentration of salt.

\subsection{Thin channel mode}\label{sec:ThinCR}

Finally, in the thin channel mode $\langle(\psi^{\prime} )^{2}\rangle$ and $\Pi$ can be approximated by~\cite{vinogradova.oi:2021}
\begin{equation}\label{eq:series_der_thin}
\langle(\psi^{\prime} )^{2}\rangle \simeq  \dfrac{4}{3 \ell_{GC}^{2}}, \, \Pi \simeq \sqrt{1 + \left( \dfrac{2 \ell_{Du}}{H}\right)^2} - 1 - \dfrac{2 \ell_{Du}}{\ell_{GC}}.
\end{equation}
Note that here we corrected a mistake in the coefficient before $\ell_{Du}$ in the expression under the square root that has been made before~\cite{vinogradova.oi:2021}.

Substituting these equations to \eqref{eq:Kfinal} yields for any $\ell_{Du}/H$
\begin{eqnarray}
  K &\simeq& K_{\infty} \left[ \sqrt{1 + \left( \dfrac{2 \ell_{Du}}{H}\right)^2} - \dfrac{4 \ell_{Du}}{3 \ell_{GC}}\left(1- \dfrac{1}{\zeta} \right)\right]\\
  &\simeq& K_{\infty} \left[\cosh \psi_s - \left(\dfrac{H}{\lambda_D}\right)^2 \dfrac{\sinh^2 \psi_s }{12}\left(1- \dfrac{1}{\zeta} \right)\right] \label{eq:hydrophilic_thinCP}
\end{eqnarray}
Here the first term is  of the leading-order, which implies that in the thin channel regime the migration should always dominate over convection.

\begin{figure}[h]
	\begin{center}
		\includegraphics[width=0.9\columnwidth]{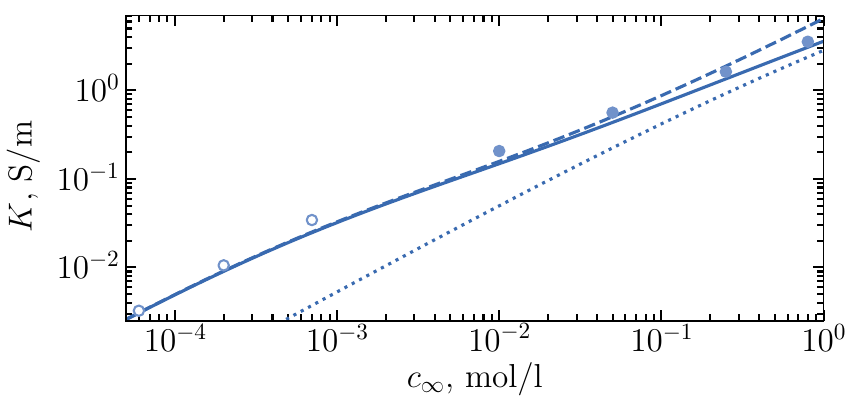}
	\end{center}
	\caption{Conductivity vs $c_{\infty}$ computed for a CP channel of $H=20$ nm, $\phi_{s}=3$ using $R=0.3$ nm. Solid curve shows calculations from Eq.~\eqref{eq:Kfinal}. Dashed curve represents $K$ calculated using ionic electrophoretic mobility from Eq.~\eqref{eq:inf}. Dotted curve corresponds bulk conductivity calculated from Eq.~\eqref{eq:Kfull}. Filled and open circles indicate calculations from Eqs.~\eqref{eq:hydrophilic_thickCP} and~\eqref{eq:hydrophilic_thinCP}.}
	\label{fig:K_cp}
\end{figure}

Figure~\ref{fig:K_cp} represents an example of the conductivity curve for a CP channel (of $H = 20$ nm). For this example we keep $\psi_s = 3$ ($\Psi_s \simeq 75$ mV) fixed and vary $c_{\infty}$. The conductivity calculations are again made only using $R = 0.3$ nm. The computed curves have been obtained following the same procedure as for curves in Fig.~\ref{fig:K_cc}. It can be seen that the conductivity strictly monotonously increases with $c_{\infty}$, and the saturation plateau does not occur. When $c_{\infty}$ is low the curves calculated with exact $\mathcal{F}$ and $\mathcal{F}=2/3$ coincide, but they separate in concentrated solutions, where exact numerical $K$ begins to approach $K_{\infty}$. We also see that the deviations from the bulk conductivity are larger for very dilute solutions. This branch of the curve corresponds to a thin channel mode, and computed data are well fitted by Eq.\eqref{eq:hydrophilic_thinCP}. In fact, with our parameters the second term can even be neglected to employ even a simpler expression $K \simeq K_{\infty}^0 \cosh \psi_s$. Thus $K$ shows the same scaling with salt as $K_{\infty}^0 \propto c_{\infty}$ being amplified in $\cosh (3) \simeq 10$ times. This situation corresponds to a ``Donnan equilibrium'' discussed above, and it would be of considerable interest to understand how the uniform ion concentrations in the channel depart from  $c_{\infty}$. The concentrations of anions and cations can be obtained from the Boltzmann distribution, $c_{\pm}/c_{\infty} = e^{\mp 3}$, which suggests that anions in the channel are significantly enriched ($c_-/c_{\infty} \simeq 20$), but cations are depleted ($c_+/c_{\infty} \simeq 0.05$). In what follows that $\cosh \psi = (c_+ + c_-)/2 c_{\infty} \simeq c_-/2 c_{\infty} \simeq 10$. Thus only anions contribute to the conductivity amplification, which is known as the ``counter-ions only'' case (or the so-called co-ion exclusion approximation) also employed before for conductivity calculations~\cite{uematsu.y:2018}. At lower $\psi_s$ the second term in \eqref{eq:hydrophilic_thinCP} comes into play and is no longer negligible. However, it always remains small.
Upon increasing $c_{\infty}$ a transition to a thick channel mode occurs. This branch of the $K$-curve is well described by Eq.\eqref{eq:hydrophilic_thickCP}. The conductivity calculations using Eq.\eqref{eq:inf} for $m$ cannot be applied at low dilution, however. It is well seen that they overestimate $K$ and result in the larger slope of the conductivity curve. Another important point to note is that with our parameters there are still discernible deviations of $K$ from $K_{\infty}$ at large concentrations, i.e. where $H/\lambda_D \gg 1$. Thus this branch of the curve  does not follow  the bulk conductivity.

\section{Concluding remarks}\label{sec:concl}

As emphasized in the introduction, this article has attempted to revisit the theory of the electrophoretic ion mobility in order to better describe the  conductivity of electrolyte solutions, both in bulk and confined in narrow channels. The main improvement of the new theory of the bulk conductivity is the extension of
its range of applicability in terms of concentrations up to a few mol/l, elimination of all arbitrary constants, enhanced accuracy, and specification of the status of prior models that are recovered in appropriate conditions.

Model~\eqref{eq:inf},  which  is  the  simplest  realistic
model for an ion mobility that one might contemplate,  provides
considerable insight  into the giant enhancement of conductivity of very dilute electrolyte solutions confined in a narrow channel. Namely, we have justified its applicability in the quite large concentration range of validity of the Laplace equation \eqref{eq:L_sphere}, where the conductivity enhancement is most pronounced. Our work clarifies that it cannot be applied at low dilution, however. In this case,  only expression \eqref{eq:M_ion_full}  can be  employed.

Our strategy can be extended to describe the case, where cations and anions of an univalent electrolyte  are  of different radius $R_+$ and $R_-$. As discussed in Appendix~\ref{a:1}, at the infinite dilution limit $R$ represents the harmonic mean of the ion radii. Although the use of the harmonic mean provided a good description of the NaCl conductivity curve up to quite large concentrations,  a detailed analysis of the conductivity at low dilution and $R_- \neq R_+$, especially when they differ significantly, remains an
open problem that would be interesting to address.

Our electrostatic mean-field model omits correlations, so the local concentration profiles do not mimic the local ordering that would occur in very concentrated and multivalent electrolyte solutions, or solvent-free ionic liquids. Further work is required to determine the systematics of the corrections to the formulas derived here for these, more complex situations. Our present model is not designed to tackle such problems, and another approach is probably better suited~\cite{bazant.mz:2011,storey.bd:2012}. However, our  present calculations should provide useful guidelines.

It would be of interest to extend the present analysis to the slippery hydrophobic surfaces~\cite{vinogradova.oi:1999}. Such surfaces can dramatically enhance an electro-osmotic flow~\cite{silkina.ef:2019} and the ionic conductivity~\cite{silkina.ef:2019,balme.s:2015} in the channel. Note that nanometric foam films that show an interesting electrokinetic behavior~\cite{bonhomme.o:2017} can also be seen as (CC) hydrophobic channels. Besides the hydrophobic slippage, the  new ``ingredient'' in such systems is the existence of a migration of adsorbed potential-determining ions in electric field~\cite{maduar.sr:2015,grosjean.b:2019}. Previous theoretical investigations of the effect of this surface migration on conductivity~\cite{mouterde.t:2018,vinogradova.oi:2021,vinogradova.oi:2022} have used Eq.\eqref{eq:inf}. The time is probably right for more detailed studies in the spirit of this paper that is by using a more appropriate expression for an ion mobility.

Finally, we recall that hydrogen and hydroxyl have been excluded from our consideration since the NLPB theory is not applicable for these ions. In fact, it remains to be seen whether the electrophoresis itself will be a good candidate to explain the high mobility of these ions. The nature of their transport is still debated: the proton-hopping (or Grotthuss) mechanism~\cite{swanson.jmj:2007} and alternative scenarios~\cite{ekimova.m:2022} have been suggested. We are unaware of any previous work that has addressed the question of entanglement of these mechanisms with the inevitable electrophoretic migration of protonated water clusters. Such an investigation appears to be a very promising direction.

\begin{acknowledgments}

  We are grateful to  G.A.Tsirlina,  who has    insistently and  diplomatically posed to us the issues about the validity of Eq.\eqref{eq:inf}. We have benefited from
discussions with E.S.Asmolov. This research was supported by the Ministry of Science and Higher Education of the Russian Federation.
\end{acknowledgments}

\appendix

\section{Validation of the theory for bulk electrolytes}\label{a:1}

\begin{figure}[h]
	\begin{center}
		\includegraphics[width=0.9\columnwidth]{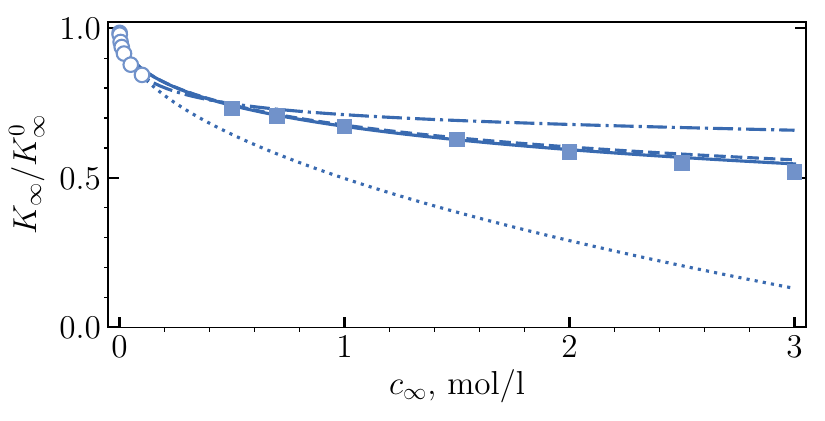}
	\end{center}
	\caption{$K_{\infty}/K_{\infty}^0$ as a function of $c_{\infty}$ for NaCl aqueous solution. Solid and dashed curves show calculations from Eq.~\eqref{eq:Kfull} and \eqref{eq:Knew2} using $R = 0.15$ nm. Circles and squares indicate experimental data from \cite{vanysek.r:2000} and \cite{lobo.vmm:1984}. Dotted and dash-dotted curves correspond to a bulk conductivity calculated from Eq.\eqref{eq:Klin} and \eqref{eq:avni}. }
	\label{fig:K_check}
\end{figure}
To check the validity of the theory, in Fig.~\ref{fig:K_check} we compare our analytical results with experimental data for NaCl in the  concentration range up to 3 mol/l~\cite{vanysek.r:2000}.  We recall that our theory was developed for ions of equal radius, but in NaCl they are of different radius, $R_+ = 0.184$ nm and $R_-= 0.125$ nm~\cite{kadhim.mj:2020}. It is straightforward to show that at the infinite dilution $R = 2 (R_+ R_-)/(R_+ + R_-)$. By this reason, for our calculations we decided to employ for $R$  the harmonic mean of hydrodynamic radii of Na$^+$ and Cl$^-$ taken from~\cite{kadhim.mj:2020}, which is equal to ca. 0.15 nm. This value seems to be too small to guarantee the high accuracy of our theory. However, the agreement of Eq.~\eqref{eq:Kfull} with experiment appears to be extremely good in the whole concentration range. Note that $\lambda_D \simeq 0.18$ nm when $c_{\infty} = 3$ mol/l, which implies that for all concentrations $\varrho = R/\lambda_D \leq 1$. This justifies the application of approximate Eq.\eqref{eq:Knew}. Using \eqref{eq:DLength} one can  rewrite \eqref{eq:Knew} as
\begin{equation}\label{eq:Knew2}
  K_{\infty} \simeq K_{\infty}^0 \left[ \dfrac{1 + 0.67 R^2 c_{\infty }}{1+ 3.28 R \sqrt{c_{\infty }}} \right],
\end{equation}
which is more convenient for analysis of experimental data. The constants in this equation are dimensional and have been calculated assuming $R$ is expressed in nm.
The calculation from \eqref{eq:Knew2} is included in Fig.~\ref{fig:K_check}. We see that it practically coincides with predictions of Eq.~\eqref{eq:Kfull}. This is a remarkable result  since Eq.\eqref{eq:Knew2} is extremely simple, easy to use, and involves only one parameter, namely, the hydrodynamic radius $R$. Also included are calculations from Eq.\eqref{eq:Klin}, which predicts a much lower
$K_{\infty}/K_{\infty}^0$ at low dilution, i.e. when $0.1 \leq \varrho \leq 1$. We emphasize that the application of the classical Onsager approach~\cite{onsager.l:1927}, which also incorporates the so-called relaxation term excluded here, would even lead to unphysical (negative) values of $K_{\infty}/K_{\infty}^0$ when $c_{\infty} \geq 2$ mol/l.

In addition, we compare our results with another calculation made with a recently published equation~\cite{avni.y:2022}, which in our notations reads:
\begin{widetext}
\centering
\begin{equation}\label{eq:avni}
  K_{\infty} \simeq K_{\infty}^0  \left[ 1 - \dfrac{R}{\lambda_D} e^{-a/\lambda_D}
    - \dfrac{1}{6} \left( 1 - \frac{1}{\sqrt{2}} + e^{-2a/\lambda_D} - \frac{1}{\sqrt{2}}  e^{-\sqrt{2}a/\lambda_D} \right)\dfrac{\ell_{B}}{\lambda_{D}} \right],
\end{equation}
\end{widetext}
where $a = 0.283$ nm is the sum of ion radii from crystallographic data. These authors infer the hydrodynamic radius from experimental data for NaCl solution  at infinite dilution~\cite{vanysek.r:2000}, which gives $R \simeq 0.15$ nm. This value coincides with the harmonic mean of hydrodynamic radii of Na$^+$ and Cl$^-$ evaluated above. The curve calculated from \eqref{eq:avni} is included in Fig.~\ref{fig:K_check}. It can be seen that at molar concentrations this equation predicts larger conductivity than the experimental data shows, and that Eq.\eqref{eq:Knew2} provides a much better fit to these data.

Note that Eq.\eqref{eq:Knew2} can be employed to infer $R$ by monitoring the relative decrease in conductivity with salt. Indeed, it follows from \eqref{eq:Knew2} that
\begin{equation}\label{eq:KnewR}
 R [\mathrm{nm}]\simeq \dfrac{2.45}{\sqrt{c_{\infty }}} \dfrac{K_{\infty}}{K_{\infty}^0} -  \dfrac{1.22}{\sqrt{c_{\infty }}} \sqrt{4.01 \left(\dfrac{K_{\infty}}{K_{\infty}^0}\right)^2 + \dfrac{K_{\infty}}{K_{\infty}^0} - 1}
\end{equation}
To verify, we take experimental data for LiClO$_{4}$ solution: $K_{\infty}/K_{\infty}^0 = 0.84$ at $c_{\infty } = 10^{-1}$ mol/l~\cite{vanysek.r:2000}. Using Eq.~\eqref{eq:KnewR} one can infer the hydrodynamic radius $R \simeq 0.2$ nm. This value is equal to a mean value of hydrodynamic radii of Li$^{+}$ and ClO$_{4}^{-}$ ions~\cite{kadhim.mj:2020}.

\section*{DATA AVAILABILITY}

The data that support the findings of this study are available within the
article.

\section*{AUTHOR DECLARATIONS}

The authors have no conflicts to disclose.


\end{document}